\documentclass[twocolumn]{jpsj2} %% two-column layout
\title{Emergence of a Barchan Belt in a Unidirectional Flow: Experiment and Numerical Simulation}

\author{Atsunari \textsc{Katsuki}$^{1,2}$\thanks{E-mail:katsuki@cp.cmc.osaka-u.ac.jp},
Macoto \textsc{Kikuchi}$^{2,1}$ and 
Noritaka \textsc{Endo}$^{3}$}

\inst{$^{1}$ Department of physics, Osaka University, Osaka 560-0043 \\
$^{2}$ Cybermedia Center, Osaka University, Osaka 560-0043 \\
$^{3}$ Department of Earth and Space Sciences, Osaka University, Osaka 560-0043}

\abst{We observed time evolution of dune fields in a water tank experiment 
and simulated it by using a simple model without taking complex fluid
dynamics into account. The initial sand bed changed its form into transverse
ripples , that is, dunes with straight crest lines perpendicular to the
flow direction. Then the crescentic shaped dunes called barchans emerged from transverse ripples.}

\kword{sand dune, barchan, transverse ripple, dune field, water tank
experiment, numerical simulation}

\begin{document}
\maketitle
%\section{Introduction} %% No sections necessary for express letters, letters and short notes
%\section{Experimental}
Sand dunes, which are found in deserts, 
on the sea bottom and also on Mars, have various morphologic patterns.
Dunes are formed by interactions
between the flow (wind or water) and sand.
 The flow makes the shape of a dune by transporting sand particles.
%transports sand particles and forms the shape of dunes.
The dune topography, in turn, acts
as a boundary condition on
the flow.
%Sand dunes have various morphologic pattern, which are found in deserts, 
%on the sea bottom and also on Mars.
%     Many types of sand dunes are found in deserts, 
%on the sea bottom and also on Mars.
%The types are found in deserts, 
%on the sea bottom and also on Mars.
Types of sand dunes are determined by the amount of available sand 
and the variation of the flow\cite{wasson1983}. 
When the flow is unidirectional 
and the amount of sand is abundant for covering the entire bedrock,
dunes with straight crest lines perpendicular to the
flow called transverse dunes are generated. 
When the flow is unidirectional but the amount of sand is insufficient, 
on the other hand,  
crescentic shaped dunes called barchans are formed. 
%dunes with straight crest lines perpendicular to the
%flow direction called transverse dunes are generated. 
Barchans usually migrate as a group and form a \textit{ barchan belt}
%, interacting 
%one another through collisions and sand flux from the
%windward
\cite{bag1941,her2004}.  
The barchan belt is generated from a sand source which exists
windward
%supposed by Bagnold
\cite{bag1941}. 
%It has been supposed that there is sand source in the windward in order
% for barchan group to exist\cite{bag1941}.
%It has been supposed that, when there is sand source in the windward,
%barchan group appear in the leeward\cite{bag1941}.
%McKee has reported that 
When there are transverse dunes windward which act 
as a sand source, 
a morphologic sequence of three types of dunes, transverse dunes, barchanoids
and barchans from upwind to downwind is observed in deserts, where barchanoid is a dune with a wavy crest
%barchans emerge from a transverse dune through a 
%barchanoid in the real desert
\cite{mckee1979}.

Observing the whole process
of the formation dynamics of a barchan belt is quite difficult in general because time scale of the dynamics is very long.
To overcome this difficulty, laboratory experiments using a water tank were proposed,
%This difficulty is overcame by a water tank experiments
\cite{nino1997,her2002,endo2004-1,endo2005,endo2006}
, which reduce the
time scale of dynamics by  less than a hundredth in rough estimates.
%Time scale of dynamics has been
%rescaled roughly below hundredth in the water tank experiments
%The water tank experiment is useful to observe the whole process
%of the formation dynamics of barchans
% because  
%the time scale of dynamics in the natural dune fields 
%has been found to be
%rescaled roughly below hundredth in the water tank experiments
%In a water tank experiment with an oscillatory flow, 
Quite recently, a transition from transverse dunes to barchans was
observed in a water tank under an oscillatory flow by Endo \textit{et. al.}\cite{endo2004-1}
A water tank experiment under an unidirectional flow
was also conducted and a similar morphologic transition was observed\cite{endo2006}.

Only a few studies on the barchan belt have been made theoretically or by
numerical simulations.
Lima \textit{et. al.}\cite{lima2002} and
Hersen \textit{et. al.}\cite{her2004} has studied 
the stability of a barchan belt,
assuming that many barchans somehow formed.
%assuming the existence of many barchans.
%assuming to be a lot of barchans already.
%Although they has assumed that the barchan belt has %existed on an initial condition,
However, formation processes of the barchan belt from a sand source have not been discussed.
%The existence of the barchan feild as an initial condition
% was assumed in these studies. 
Moreover their models have not taken into account 
a solitary wave behavior of barchans, which was 
found in collision dynamics\cite{schw2003-3,katsuki2004-1,duran2004}.
% was not taken into account. 
In this letter, we will 
investigate formation dynamics of a barchan belt by using a numerical model, 
which has succeeded in reproducing a solitary wave behavior\cite{katsuki2004-1}.
We also conduct a water tank experiment 
for the purpose of observing development of a barchan belt.
\begin{figure}
\begin{center}
\includegraphics[width=1.\linewidth]{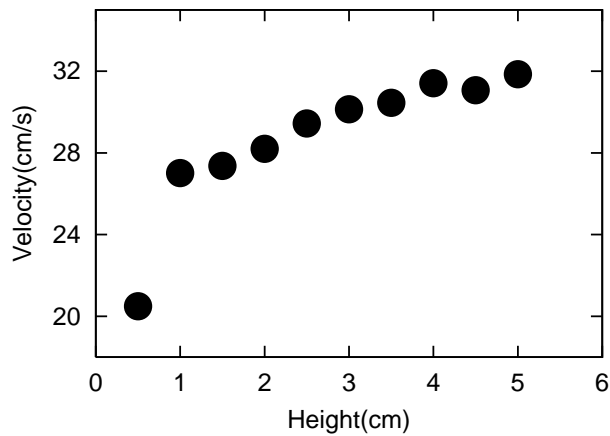}
%\end{center}
 \caption{Height dependence of the flow velocity in a water tank experiment. 
%The sand bed on the initial conditions is
% formed into transverse ripples. After a while the transverse ripples generate
% barchan dunes.
}
\end{center}
\label{fig:flow}
\end{figure}
%The time scale of dynamics in the natural dune fields has been found to be
%rescaled 
%roughly below hundredth in the water tank experiments.
\begin{figure}
\begin{center}
\includegraphics[width=1.\linewidth]{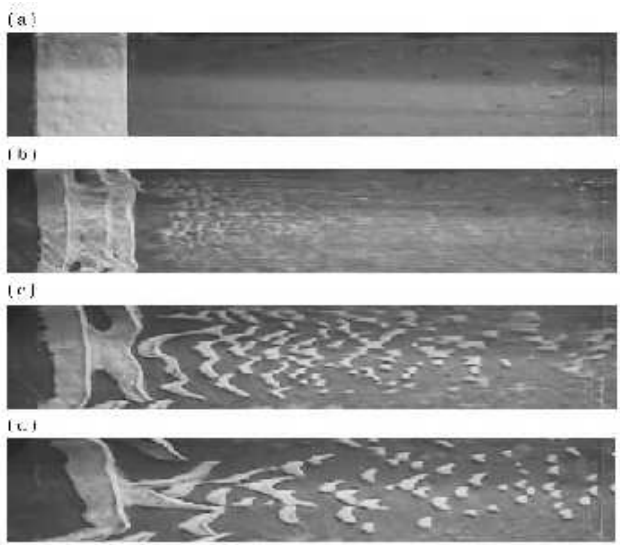}
\end{center}
 \caption{Time evolution of a dune field in the water tank experiments
under a steady unidirectional flow. (a) 0 second (initial configuration), (b) 180 seconds, (c) 480 seconds
 and (d)1140 seconds.
%The sand bed on the initial conditions is
% formed into transverse ripples. After a while the transverse ripples generate
% barchan dunes.
}
\label{fig:f1}
\end{figure}
%Thus, the experiments enable us to observe the whole processes of dune
%field dynamics.
%In this letter, we will 
%provide a figure of dune feild with transition processes in a water
%tank experiment under an unidirectional flow in order to compare with a
%numerical simulation and  
%investigate formation dynamics of a barchan belt 
%what is the main process of barchan group 
%by a numerical simulation.
%We also conduct a water tank expeiment.
%Besides, a picture of dune feild with transition processes 
%show a barchan group with transition processes 
%investigate how the barchan group form by
%will be provided 
%in a water
%tank experiment under an unidirectional flow in order to compare with a %numerical simulation.
%and investigate what is the main process of barchan group by a
%numerical simulation. 
%These sand dunes have
%been investigated by a lot of geomorphologists and physicists.
%The interaction between the flow (wind or water) and sand forms dunes.
%The flow transports sand particles 
%and forms the shape of dunes.
%The dune topography, in turn, acts
%as a boundary condition on
%the flow.
%\cite{bag41,Pye1990,cooke1993,lancaster1995}.
%\cite{bag41}
% since R.A.Bagnolds studied fifty years ago \cite{bag41}. 

%A water t
%from a relationship between flow velocity and
%height, where data was taken in a run without a sand bed%(Fig.\ref{fig:flow}).
 \begin{figure}[t]
\begin{center}
\includegraphics[width=0.7\linewidth]{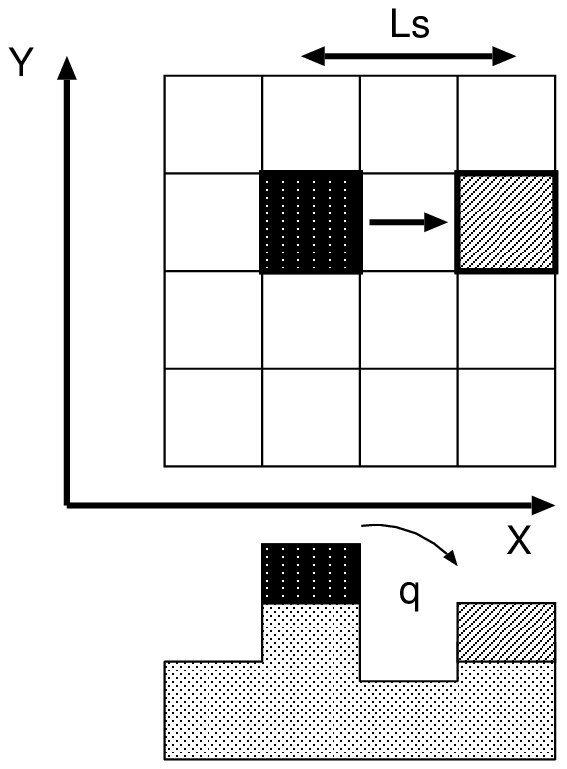}
\end{center}
 \caption{Schematic diagram of saltation process. $L_S$ and q are saltation
 length and saltation mass, respectively.
%The sand bed on the initial conditions is
% formed into transverse ripples. After a while the transverse ripples generate
% barchan dunes.
}
\label{fig:sal-zu}
\begin{center}
\includegraphics[width=0.8\linewidth]{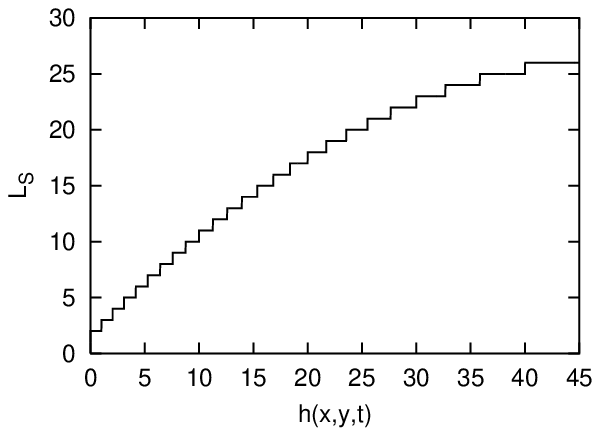}
	\caption{\label{fig:salkaidan.eps}Relationship between the
 saltation length and the dune height (eq.(\ref{eq:Ql})). Saltation length is
 rounded up to an integer.}
\end{center}
\end{figure}
%\begin{figure}
%\begin{center}
%\includegraphics[width=0.8\linewidth]{./eps/salkaidan.eps}
%	\caption{\label{fig:salkaidan.eps}Relationship between the
% saltation length and the dune height (eq.(\ref{eq:Ql})). Saltation length is
% rounded up to an integer.}
%\end{center}
%\end{figure}

The experiment was performed in a  water tank 
%which was 
(10 m long, 20 cm wide and 50 cm deep) 
in which
a pump can generate a steady unidirectional flow. 
Water depth 13.5 cm was kept constant.
The sand was well-sorted: 80-100 $\mu m$ in diameter and 130 g in total
weight. 
%The flow velocity was 28 cm/s at 1 cm above the flume bottom. 
%Figure \ref{fig:flow} showed a relationship between flow velocity and
%height, where data was taken in a run without a sand bed.
%The boundary layer was supposed to be about 1 cm.  
In the initial condition, the sand bed was placed in an area of 15 cm long
and 20 cm wide.
Figure \ref{fig:flow} shows the flow velocity measured at several different heights in a run without a sand bed.
Since height of the sand bed was less than 1 cm,
 the transportation of 
sand was considered to occur inside the boundary layer. 

After the flow started, 
stages of morphologic transition were
observed.
% The sand bed changed, through transverse ripples,
%to barchan dunes. 
%When the sand bed changed to transverse ripples, 
First, instability on the sand bed grew and transverse ripples started
to appear
(Fig. \ref{fig:f1}(b)).
Next, parts of the transverse ripples were breached at some places and the
fragments eventually became barchans (Fig. \ref{fig:f1}(c) and \ref{fig:f1}(d) ).
Figure \ref{fig:f1}(d) closely resembles a barchan belt in a real desert sketched by Bagnold\cite{bag1941}.

%    We confirmed the formation dynamics of a barchan group in
%above.
% Next,  main processes of dune feild dynamics
%formation dynamics of a barchan group
%will be investigated  by a numerical simulation.
%a numerical simulation of it will be performed in order to
%investigate main processes.
%In general, dunes is formed the interaction
%between the flow (wind or water) and sand.
% The flow transports sand particles 
%and forms the shape of dunes.
%The dune topography, in turn, acts
%as a boundary condition on
%the flow.
%     Numerical simulation of a dune field including processes of flow and sand% is ordinarily 
%a formidable task because of
%computational cost. In addition, 
%if complex process as flow dynamics and granular
% dynamics were taken into account in detail, 
%the complexity might disturb to find
%intrinsic processes.   
  Numerical simulation of a dune field taking into account  complex process of
 flow dynamics and granular dynamics is
a formidable task because of  computational cost. 
%the complexity might disturb to find
%intrinsic processes.   
%In addition, it is possible for the complexity to cover intrinsic processes.   
%In addition, 
%intrinsic processes which govern the dynamics are covered by other factor
%and  
%In order to reduce the cost,
In what follows, we used a simple model 
with minimal processes.
Both complex fluid dynamical processes
such as convection and viscosity and effect of size and shape of sand grain were ignored.
%but having minimal processes only.
This model has already succeeded in
reproducing morphologic features of a single barchan and collision dynamics of
two barchans\cite{katsuki2004-1}.

     In the model,
the dune field  is divided into square cells.
%was constructed on cells discretely
\cite{werner1995,nishi1998}
Each cell is considered to represent an 
area of the ground which is sufficiently larger than the sand grains. 
%Each cell was considered to represent an 
%The unit of a position $(x,y)$ was denoted as a lattice %length.
% Each scale of each lattice was an 
%area of the ground which was sufficiently larger than the sand grains. 
%instead of calculating a motion of each sand grain 
%because of reducing calculation cost largely.
%Also time step $t$ in numerical simulation was denoted discrete variables.
A field variable $h(x,y,t)$ which expresses the local surface height
is assigned to each cell; $t$ denotes the discrete time step and
the spatial coordinate $x$ and $y$ are positions  of the center of a cell in the flow direction and in the lateral direction, respectively.
The edge length of the cell is taken as a unit of length.
In short,  $x$, $y$ and $t$ are 
discrete variables while $h(x,y,t)$ takes a
continuous value.

Next, we models a motion of sand grain considering only the following two processes: saltation and avalanche.
%only two processes, saltation and avalanche, was used in this work in
%order to simplify a model.
Saltation is the transportation process of sand grains by the flow.
The schematic diagram of the saltation is shown in
 Fig. \ref{fig:sal-zu}. 
%The length 
%transported by saltation and a sand mass were called
The saltation length and saltation mass are denoted  $L_S$ and $q$, respectively.
Here, the saltation mass indeed is a volume of sand transfered from a cell to another cell.
Since the area of a cell is unity, $q$ represents change of height of a cell by saltation.
In each time step of a simulation, the saltation mass $q$
% which was fixed as 0.1 in order to simplify this model, 
shifts  
from a cell $(x,y)$ to the leeward cell $(x+L_S,y)$.
%, where $L_S$ was
%rounded up due to be integer. 
Hence changes of height $h(x,y) \rightarrow  h(x,y)-q$ and $h(x+L_S,y)
\rightarrow  h(x+L_S,y)+q$ take place at the taking-off cell and the 
landing cell, respectively.
Saltation length
is then modeled by the following equation:  
\begin{eqnarray}
L_S &=& a+bh(x,y,t)-ch^2(x,y,t).
%q &=& d.
\label{eq:Ql}
\end{eqnarray}
 The phenomenological 
parameters $a$=1.0, $b$=1.0 and $c$=0.01 are used in this work.
$L_S$ is
rounded up  to an integer as is shown in Fig. \ref{fig:salkaidan.eps}
. 
Equation (1) represents that the sand is transported farther away as
height is higher, but not too far.   
Note that equation (1) is used only in the range in which $L_S$ is 
an increasing function of $h(x,y,t)$.
The saltation mass $q$ is fixed as 0.1 for simplicity throughout the present work.

Avalanche, on the other hand, is the process in which the sand slides down along the steepest slope if it becomes steeper than the angle of repose.
%until the slope relaxed to be at (or lower than)
%the angle of repose, 
%if the angle of slope exceeded it.
% The angle of repose was defined as $34^{\circ}$.
% in this simulation.
In this simulation, first, the cells are marked if the slope, which 
makes with the lowest of their nearest-neighbor 
cells, exceeds the angle of repose.
Then half of the excess volume of each marked cell is transfered.
Here the angle of repose is fixed as $34^{\circ}$.
%Sand slid down a cell $(x_NNA,y_NNA,t)$ which has the most difference of height %$\delta h =h(x,y,t)-h(x_NN,y_NN,t)$ in the nearest neighbor cells $h%(x_NN,y_NN,t)$ of $(x,y)$, if the local slope was over the angle of repose.
%The amount of sand $Q_A$ for avalanche process in a iteration was
%$Q_A=1/2(h(x_NN,y_NN,t)-h(x,y,t)-tan 34^{\circ})$ where 1/2 means the height of %cell before the sand transported was not over after the sand transports.
%The height was changed $h(x,y,t) \rightarrow  h(x,y,t) -Q_A$ and $h%(x_NNA,y_NNA,t) \rightarrow  h(x_NNA,y_NNA,t) +Q_A$.
This procedure is repeated until all cells satisfy the stability condition.
Two processes of saltation and avalanche are performed by turns.
% and one time step was determined as one turn.
%Note that the unit of height and the length of a cell were related to
%each other only via the angle of repose. 

\begin{figure}
\begin{center}
\includegraphics[width=0.8\linewidth]{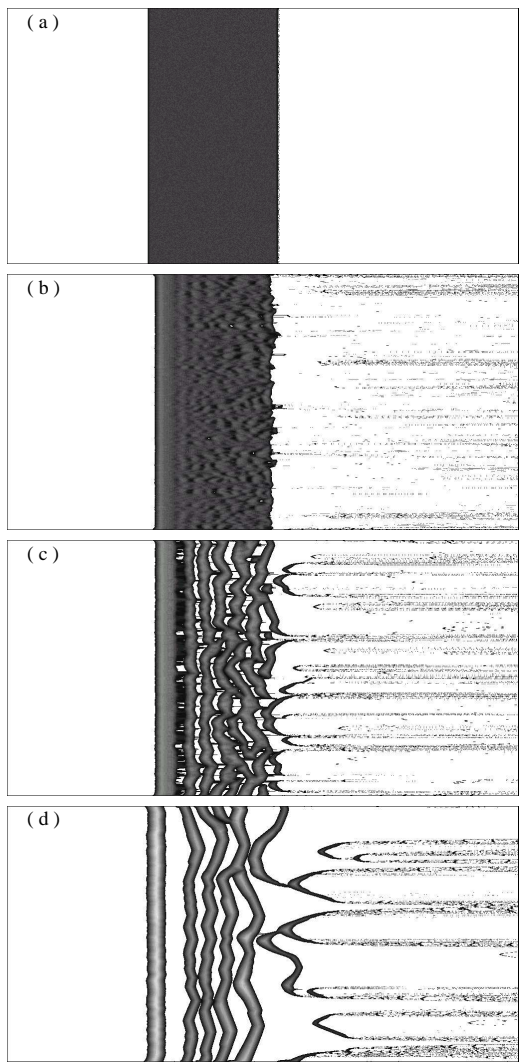}
\end{center}
\caption{Time evolution of the dune field for $D_0 = 3$ and $L_0 =256$  in
 numerical simulation. 
%Sand bed with random noise is set
% on the initial condition. The sand bed is
% formed into transverse ripples. Moreover the transverse ripples generate
% barchan dunes.
The center of mass coordinate is
fixed in the figure. Time steps are (b) 1000, (c) 2000 and (d) 5000.}
\label{fig:f3}
\end{figure}

\begin{figure}[h]
\begin{center}
\includegraphics[width=0.7\linewidth]{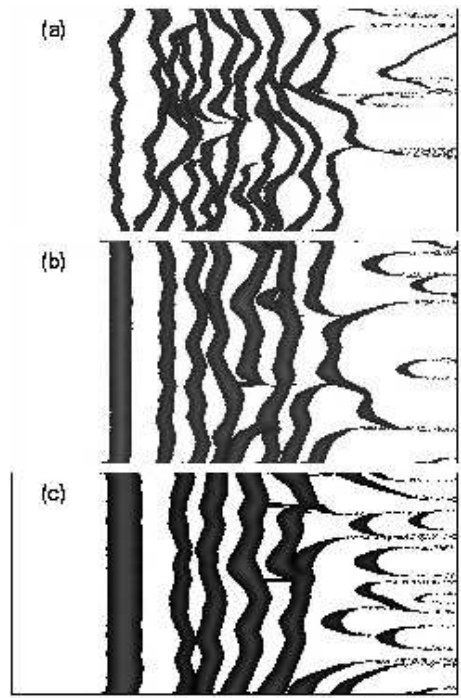}
\end{center}
\caption{Dependence of ripple size on the depth $D_0$.
 The length and
  time steps were fixed as $L_0 = 512$ and $t = 10000$, respectively. 
% (a) $D_0 = 2 $ and (b) $D_0 = 6 $.}
 (a) $D_0 = 2 $, (b) $D_0 = 4 $ and (c) $D_0 = 6 $.}
\label{fig:f4}
\end{figure}
\begin{figure}
\begin{center}
\includegraphics{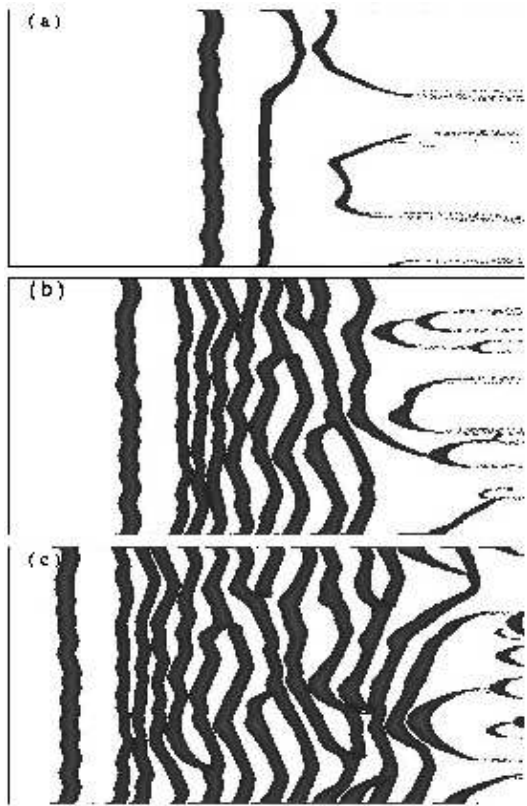}
\end{center}
\caption{Dependence of ripple size on the length $L_0$. The depth and
  time steps were fixed as
 $D_0 = 3$ and $t = 10000$, respectively. 
 (a) $L_0 = 128 $, (b) $L_0 = 512 $ and (c) $L_0 =  768$.}
\label{fig:f5}
\end{figure}
%\begin{figure}
%\begin{center}
%\includegraphics{fig5.eps}
%\end{center}
%\caption{Dependence of ripple size on the length $L_0$. The depth and
%  time steps were fixed as
% $D_0 = 3$ and $t = 10000$, respectively. 
% (a) $L_0 = 128 $, (b) $L_0 = 512 $ and (c) $L_0 =  768$.}
%\label{fig:f5}
%\end{figure}
 In the numerical field,
the open
boundary condition was imposed in the flow direction while the lateral
boundary was impenetrable to sand grains.
It means that there was no influx sand from the upwind boundary and 
sand which went out from the downwind boundary was simply deleted.
%The  dune field consisted of 1024 (windward) by 512 (lateral) cells.
The dune field consisted of 1052 (windward) by 512 (lateral) cells.
%The sand are accumulated near walls in water tank experiment. 

     As an initial condition, we set a rectangular sand bed in order to
reproduce the experimental situation. 
The initial sand bed had a sand depth $h(x,y,0) = D_0 = 3$, 
length $L_0 =256 $ along the flow direction and width
$W_0 = 512 $ (Fig. \ref{fig:f3}(a)).
%The initial sand bed had a sand depth $h(x,y,0) = D_0 = 3$, 
%length $L_0 = 256 $ along the flow direction and width
%$W_0 = 512$ (Fig. \ref{fig:f3}(a)).
Height of the sand $h(x,y,t)$ were then
perturbed with
 $\pm$ 10 $ \verb+%+ $
random noise.  
%Figure.1(a) shows a sand depth D=h(x,y,0)=3,
% which correspond to a height as length of a cell is 1, 
%and a portion of length L=256/1024=0.25 of the initial sand bed.

 After the simulation started,
%the sand bed changed to transverse ripples
%because of instability in the flow direction.
instability grew in the flow direction and
%instability in the flow direction 
made the sand bed change to transverse ripples.
%instability grew in the sand bed, and the sand bed was formed into transverse
%ripples.
%sand on the sand bed are 
%transported here and there . 
%After a while,
% the sand bed becomes transverse ripples.
%When more times go,
After a while,
%a part of transverse ripples is piled by sand and 
barchans emerged from the transverse ripples
because of instability in the lateral direction (Fig. \ref{fig:f3}) .
%by the effect of instability in the lateral direction (Fig. \ref{fig:f3}).
%instability in the lateral direction 
The processes quite similar to what were observed in the water tank
experiment were thus reproduced. 
%This pattern transition from starved ripples to barchans has hardly
%discussed in simulation because of taking a calculation cost largely.

     Next, we changed the sand depth  $D_0$ and the length $L_0$ of the
     initial sand bed.
% where the dune field consisted of 1024 (windward)
%     by 512 (lateral) cells.
%We found that 
As $D_0$  became shallower,
ripples were found to be narrower (Fig. \ref{fig:f4}).
% The number of the ripples, on the other hand, increased
%as  $L_0$  was larger (Fig. \ref{fig:f5}).
As $L_0$  became larger,  on the other hand, the width of ripples was 
unchanged but the number of ripples increased (Fig. \ref{fig:f5}). 
It means that the ripples have a characteristic length which depends on
initial depth of the sand bed as long as $D_0$ is much larger than $q$. 

\begin{figure}[t]
\begin{center}
\includegraphics[width=1.\linewidth]{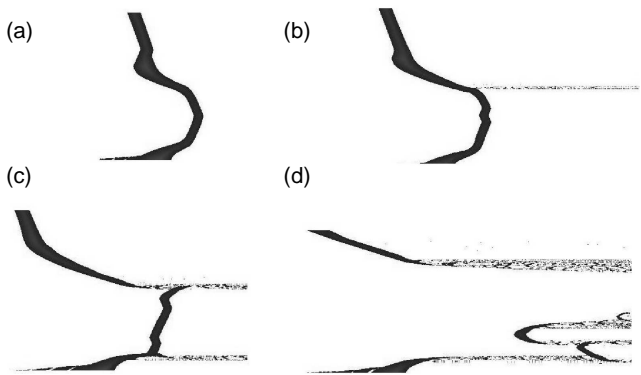}
\end{center}
\caption{
%Configuration between a barchan and fragments of
%transverse ripple after the breaching.
Type A transition from a transverse ripple to a barchan.
The barchan appeared in the  leeward  side from fragments of transverse
 ripple. Initial condition for sand bed are $D_0 = 2.5$ and $L_0 =256$. Time
 steps are (a) 45000, (b) 50000, (c) 55000 and (d) 65000.}
%Before a barchan emerged from the transverse ripple,
%a part of sand often escaped from 
%transverse ripple.}
\label{fig:f5.5}
\end{figure}

\begin{figure}[t]
\begin{center}
\includegraphics[width=1.0\linewidth]{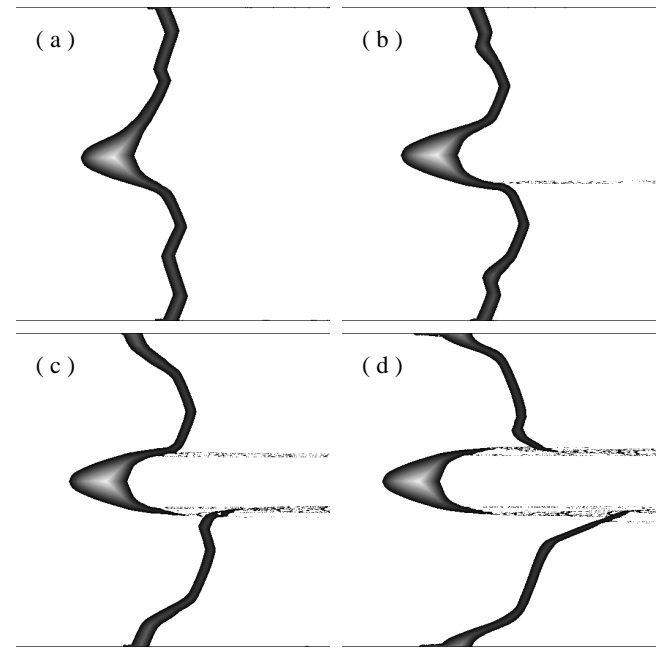}
\end{center}
\caption{
Type B transition from a transverse ripple to a barchan.
The barchan appeared in the  windward  side from fragments of transverse
 ripple. Initial condition for sand bed are $D_0 = 3$ and $L_0 =64$. Time
 steps are (a) 65000, (b) 70000, (c) 75000 and (d) 80000.}
%Configuration between a barchan and fragments of
%transverse ripple after the breaching.
%There was a barchan in the windward side from fragments of transverse
% ripple. On the initial condition for $D_0 = 3$ and $L_0 =64$, time
% steps were (a) 65000, (b) 70000, (c) 75000 and (d) 80000.}
%Before a barchan emerged from the transverse ripple,
%a part of sand often escaped from 
%transverse ripple.}
\label{fig:f6}
\end{figure}

We observed two types of transition from transverse ripples to barchans.
Type A, an example of which is shown in Fig. \ref{fig:f5.5}, occured 
when a portion of a transverse ripple happened to be much lower than the rest body. The transverse ripple gradually extended to the leeward at that part and 
eventually breached because the migration velociy is inversely proportional to the height\cite{finkel1959,hesp1998,andre2002-1}.
After that, the fragment thus formed became a barchan.
Transition of this type were also observed 
in the water tank experiment \cite{endo2006}.
%In one type, there was a barchan in the leeward side from fragments of
%transverse ripple (Fig. \ref{fig:f5.5}).
%This type was observed when the initial sand bed was comparatively deep.
%This type was observed when 
%There was a case that 
% portion of transverse ripple happened to be lower than the other portion of
% the same transverse ripple. Then, a portion of
%transverse ripple was extended to the leeward side
%gradually and was breached,  
%because dune hold the roughly inverse relation between migration velocity 
%and height\cite{finkel1959,hesp1998,andre2002-1}.
%This configuration of a barchan in the leeward side from fragments of
%transverse ripple was formed.
%This configuration was a barchan in the leeward side from fragments of
%transverse ripple
%, a portion of
%transverse ripple was extended to the leeward side
%gradually and was breached. Then the breached dune became a barchan.
%A portion of transverse ripple was extended to the leeward side
%gradually and was breached. Then the breached dune became a barchan
%This configuration was observed by experiment\cite{endo2006}.
Type B, an exapmle of which is shown in Fig. \ref{fig:f6}, occured
when a portion of a transverse ripple happened to be much higher than the rest body. The transverse ripple formed a crescent shape at that part, 
which evetually breached from a ripple to form an isolated barchan.
In this case, a barchan appeared in the windward of the ripple.
%The transverse ripple went away faster to the downwind than the portion of the %crescent shape and breached.  
%In another type, there was a barchan in the windward side from fragments of
%transverse ripple (Fig. \ref{fig:f6}).
%This type was observed when the initial sand bed was comparatively
%shallow.
%In this type was observed when
%a portion of transverse ripple happen to be higher than the other portion of
% the same transverse ripple. The higher portion changed into the shape
% of barchan. Since the velocity of the barchan became slower than the ripple
% except a portion of barchan, the ripple went away faster
% to the leeward and This configuration was made.
This type was observed only by the simulation so far.
Whether or not this type can be observed in a water tank is left for a future work.
%After a while, the ripple except a portion of barchan went away
% to the leeward 
%because of the difference of velocity between barchan and the ripple. 
%After a while, the shape of barchan at a transverse ripple was formed 
%and the ripple except a portion of barchan went away to the leeward 
%because of the difference of velocity between barchan and the ripple.    
%This results are important when observer will deduce in environment condition
%from a dune configuration.
 
%     Next, we investigated motion of the 
%transported sand in detail in order to
%understand how barchans emerged.
 We found that in most cases, although not always, a portion of the sand 
escaped from a transverse ripple before a barchan emerged.
% Figure \ref{fig:f6} showed an example  
%for $D_0 = 3 $  and $L_0 = 64 $. 
%It is difficult to observe the sand escape in the water tank experiments
%because 
%This sand escape can also be seen in the water tank experiments 
%, although it is not clear if emergence of
% a barchan is always accompanied by the sand escape 
%because observing sand escape the amount of which is a little is difficult.
% unless the amount is large to some extent.
This sand escape can also be seen in the water tank experiments,
 although it is not clear whether or not emergence of a barchan is always 
accompanied by the sand escape, because observing it 
is difficult unless the amount 
is large enough.
%difficult to observe it clearly. 
%Although a barchan did not always
%emerge even if a part of the sand escaped from a transverse ripple, 
%it will be advantageous to focus on the area where the sand escape from
%a ripple for observing the emergence of the baarchan in the natural dune
%fields and experiments. 
%A barchan did not always
%emerge even if some sand escaped from a transverse dune.
% sand escape is conceivable as a prediction of emergence of barchan.
%It means that sand escape was symptom of barchans formation.
%not a result of formation of barchan.
The sand escape is 
a precursor of a barchan formation, so that it 
%The sand escape phenomena 
will be useful for locating the place where 
a barchan is emerging in natural dune fields.
%It is sometimes observed that a barchan fails to emerge even if some sand
%escaped from a transverse dune.
%The sand escape is 
%a precursor of a barchan formation.

%This phenomena of sand escape is useful for observing the emergence of %barchans in natural dune
%field. 

     In conclusion, we succeeded in simulating a dune field dynamics 
by using a simple model in a steady unidirectional flow.
%an experiment of a dune field dynamics in a
%water tank and to simulate it
% by using the simple model in a steady unidirectional flow.
The model took only two processes into account: saltation and
 avalanche.
 The barchans emerged through transverse ripples from the initial sand
bed. Before barchans formed from the transverse ripples, a portion of
sand often escaped from the ripples. 
%Besides,  the length of initial sand bed are larger to the windward direction, the number of transverse ripples is more. Moreover as the depth of initial sand bed is deeper, the height of transverse dune is higher. 

We thank to H. Nishimori for useful discussion and K. Taniguchi for
     experimental support.
     This work was partially supported by 
the 21st Century COE Program named "Towards a new basic science : 
 depth and synthesis".

%\bibliography{note05}
%\section{Discussion}
%
%\subsection{Subsection sample 1}

\end{document}